\documentclass{eptcs}
\providecommand{\event}{FICS 2012} 
\usepackage{breakurl}              

\usepackage{proof}

\usepackage{newlfont}

\usepackage{amssymb}
\usepackage{latexsym}

\usepackage[all]{xy}

\usepackage{graphicx}
\usepackage{color}

\newcommand{\al}{\alpha}
\newcommand{\be}{\beta}

\newcommand{\pair}[2]{\langle #1, #2\rangle}

\newcommand{\List}{\mathsf{List}}
\newcommand{\El}{\mathsf{El}}
\newcommand{\nil}{\mathsf{nil}}
\newcommand{\cons}{\mathsf{cons}}
\newcommand{\isort}{\mathsf{isort}}
\newcommand{\ins}{\mathsf{ins}}
\newcommand{\qsort}{\mathsf{qsort}}

\newcommand{\qsplit}{\mathsf{qsplit}}

\newcommand{\Str}{\mathsf{Str}}
\newcommand{\hd}{\mathsf{hd}}
\newcommand{\tl}{\mathsf{tl}}
\newcommand{\smerge}{\mathsf{smerge}}
\newcommand{\dropevens}{\mathsf{dropeven}}

\newcommand{\red}[1]{#1}

\newcommand{\Dom}{\mathsf{dom}}
\newcommand{\Domg}{\mathsf{Dom}}
\newcommand{\Domc}{{\mathsf{Dom}^\infty}}

\newcommand{\dn}{\mathbin{\downarrow}}
\newcommand{\dnc}{\mathbin{\downarrow^\infty}}

\newcommand{\inl}{\mathsf{inl}}
\newcommand{\inr}{\mathsf{inr}}

\newcommand{\fst}{\mathsf{fst}}
\newcommand{\snd}{\mathsf{snd}}

\title{Structured general corecursion and coinductive graphs [extended abstract]}
\author{Tarmo Uustalu
\institute{Institute of Cybernetics at Tallinn University of Technology, Estonia}
\email{tarmo@cs.ioc.ee}
}
\def\titlerunning{Structured general corecursion and coinductive graphs}
\def\authorrunning{T.~Uustalu}

\begin{document}

\maketitle

\begin{abstract}
  Bove and Capretta's popular method for justifying function
  definitions by general recursive equations is based on the
  observation that any structured general recursion equation defines
  an inductive subset of the intended domain (the ``domain of
  definedness'') for which the equation has a unique solution. To
  accept the definition, it is hence enough to prove that this subset
  contains the whole intended domain.

  This approach works very well for ``terminating'' definitions. But
  it fails to account for ``productive'' definitions, such as typical
  definitions of stream-valued functions. We argue that such
  definitions can be treated in a similar spirit, proceeding from a
  different unique solvability criterion. Any structured recursive
  equation defines a coinductive relation between the intended domain
  and intended codomain (the ``coinductive graph'').  This relation in
  turn determines a subset of the intended domain and a quotient of
  the intended codomain with the property that the equation is
  uniquely solved for the subset and quotient. The equation is
  therefore guaranteed to have a unique solution for the intended
  domain and intended codomain whenever the subset is the full set and
  the quotient is by equality. 
\end{abstract}

\paragraph{Unique solutions to recursive equations}

General recursive definitions are commonplace in programming practice.

In particular, it is highly desirable to be able to define functions
by some forms of controlled general recursion in type-theoretically
motivated languages of total functional programming (in particular,
proof assistants) that come with a set-theoretic rather than a
domain-theoretic semantics. For an overview of this area, see Bove et
al. \cite{BKS:parrit}.

In this paper, we are concerned with describing a function $f : A \to
B$ definitely by an equation of the form:
\begin{equation}
\label{eq:rec}
\xymatrix{
FA \ar[d]_{F f}
  & A \ar[l]_-{\al} \ar[d]^{f} \\
FB \ar[r]_-{\be} 
  & B
} 
\end{equation}
where $A$, $B$ are sets (the intended domain and codomain), $F$ is a
functor (the branching type of recursive call [corecursive return]
trees), $\al$ is an $F$-coalgebra structure on $A$ (marshals arguments
for recursive calls) and $\be$ is an $F$-algebra structure on $B$
(collects recursive call results). We are interested in conditions
under which the equation is guaranteed to have a unique solution
(rather than a least solution in a domain-theoretic setting or some
solution that is canonical in some sense). There are several important
generalizations of this setting, but we will not treat them here.

There are some well-known good cases. 

\paragraph{Some good cases (1): Initial algebra}

The following equation has a unique solution for any $B$, $\be$.
\[
\xymatrix@C=6em{
1 + \El \times \List \ar[d]_{1 + \El \times f}
  & \List \ar[l]_-{\red{[\nil,cons]^{-1}}} \ar[d]^{f} \\
1 + \El \times B \ar[r]_-{\be} 
  & B
}
\]
E.g., for $B = \List$ (lists over $\El$), $\be = \ins$ (insertion of an
element into a list assumed to be sorted), we get $f = \isort$
(insertion sort).

A unique $f$ exists because $(\List, [\nil,\cons])$ is the
\emph{initial algebra} for the functor $F X = 1 + \El \times X$. It is
the \emph{fold} (the unique algebra map) determined by the algebra
$(B, \be)$.

\paragraph{Some good cases (2): Recursive coalgebras}

A unique solution exists for any $B$, $\be$ also for the equation
\[
\xymatrix@C=6em{
1 + \El \times \List \times \List \ar[d]_{1 + \El \times f \times f}
  & \List \ar[l]_-{\red{\qsplit}} \ar[d]^{f} \\
1 + \El \times B \times B \ar[r]_-{\be} 
  & B
}
\]
where $\qsplit\, \nil = \inl\, *$ and $\qsplit\, (\cons\, ( x, xs)) =
\inr\, (x,  xs|_{\leq x}, xs|_{> x})$.
E.g., for $B = \List$, $\be = \mathsf{concat}$ (concatenation of the
first list, the element and the second list), we get $f = \qsort$
(quicksort).

$(\List, \qsplit)$ is \underline{not} the inverse of the initial
algebra of $F X = 1 + \El \times X \times X$ (which is the algebra of
binary node-labelled trees), but we still have a unique $f$ for any
$(B, \beta)$.

For this property, $(\List, \qsplit)$ is called a \emph{recursive
  coalgebra} of $F$.  Recursive $F$-coalgebras form a full subcategory
of the category of all $F$-coalgebras.  The inverse of the initial
$F$-algebra is the final recursive $F$-coalgebra.

While recursiveness is a very useful property of a coalgebra, it is
generally difficult to determine whether a given coalgebra is
recursive. For more information on recursive coalgebras, see Taylor
\cite{Tay:prafm}, Capretta et al. \cite{CUV:reccc}, Ad\'amek et
al.~\cite{ALM:reccff}.

\paragraph{Some good cases (3): Final coalgebra}

This equation has a unique solution for any $A$, $\al$.
\[
\xymatrix@C=6em{
\El \times A \ar[d]_{1 + \El \times f}
  & A \ar[l]_-{\al} \ar[d]^{f} \\
\El \times \Str \ar[r]_-{\red{\pair{\hd}{\tl}^{-1}}} 
  & \Str
}
\]
E.g., for $A = \Str$ (streams), $\al = \pair{\hd}{\tl \circ \tl}$ (the
analysis of a stream into its head and the tail of its tail), we get
$f = \mathsf{dropeven}$ (the function dropping every even-position
element of a given stream).

A unique $f$ exists for any $(A, \alpha)$ because $(\Str,
\pair{\hd}{\tl})$ is the \emph{final coalgebra} of $F X = \El \times
X$. It is the \emph{unfold} (the unique $F$-coalgebra map) given by
the coalgebra $(A, \al)$.

\paragraph{Some good cases (4): Corecursive algebras 
}

This equation has a unique solution for any $A$, $\al$:
\[
\xymatrix@C=6em{
\El \times A \times A \ar[d]_{\El \times f \times f}
  & A \ar[l]_-{\al} \ar[d]^{f} \\
\El \times \Str \times \Str \ar[r]_-{\red{\smerge}} 
  & \Str
}
\]
Here $\hd\, (\smerge (x, xs_0, xs_1)) = x$ and $\tl\, (\smerge (x, xs_0,
xs_1)) = \smerge (\hd\, xs_0, xs_1, \tl\, xs_0)$.

$(\Str, \smerge)$ is not the inverse of the final coalgebra of $F X =
\El \times X \times X$, but a unique $f$ still exists for any $(A,
\alpha)$. We say that $(\Str, \smerge)$ is a \emph{corecursive
  algebra} of $F$, cf. Capretta et al.~\cite{CUV:corasg}. [The inverse
of the final $F$-coalgebra is the initial corecursive $F$-algebra and
thus a special case.]  Similarly to recursiveness of a coalgebra,
corecursiveness of an algebra is a useful property, but generally
difficult to establish.

\medskip

The equation~\ref{eq:rec} can of course have a unique solution also in
other cases. In particular, it may well happen that neither is $(A,
\al)$ corecursive nor is $(B, \be)$ recursive, but the equation still
has exactly one solution.

\paragraph{General case (1): Inductive domain predicate} 

Bove and Capretta \cite{BC:modgrt,BC:typprf} put forward the following
approach to recursive definitions in type theory (the idea has
occurred in different guises in multiple places; it must go back to
McCarthy): for a given recursive definition, work out its ``domain of
definition'' and see if it contains the intended domain.

For given $(A, \alpha)$, define a predicate $\Dom$ on $A$
\emph{inductively} by
\[
\infer{\Dom\, a}{
  a : A 
  &
  (\tilde{F}\, \Dom)\, (\al\, a)
}
\]
(i.e., as the smallest/strongest predicate validating this rule),
denoting by $\tilde{F}\, P$ the lifting of a predicate $P$ from $A$ to
$F\, A$.

Write $A|_\Dom$ for the subset of $A$ determined by the predicate
$\Dom$, the ``domain of definedness''. It is easily verified that, for
any $(B, \be)$, there is $f : A|_\Dom \to B$ uniquely solving
\[
\xymatrix{
F(A|_\Dom) \ar[d]_{F f}
  & A|_\Dom \ar[l]_-{\al|_\Dom} \ar[d]^{f} \\
FB \ar[r]_-{\be} 
  & B
}
\]
If $\forall a : A.\, \Dom\, a$, which is the same as $A|_\Dom \cong
A$, then $f$ is a unique solution of the original
equation~\ref{eq:rec}, i.e., the coalgebra $(A, \al)$ is recursive.

For $A = \List$, $\al = \qsplit$, $\Dom$ is defined inductively by
\[
\infer{\Dom\, \nil}{
}
\qquad 
\infer{\Dom\, (\cons\, (x, xs))}{
  x : \El
  & 
  xs : \List
  & 
  \Dom\, (xs|_{\leq x})
  & 
  \Dom\, (xs|_{> x})
}
\]
We can prove that $\forall xs : \List.\, \Dom\, xs$. Hence $(\List,
\qsplit)$ is recursive.

If $A|_\Dom \cong A$, the coalgebra $(A, \al)$ is said to be
\emph{wellfounded}. Wellfoundedness gives an induction principle on
$A$: For any predicate $P$ on $A$, we have
\[
\infer{P\, a}{
  a : A
  &
  \infer*{P\, a'}{
    a' : A 
    &
    (\tilde{F}\, P)\, (\al\, a')
  }
}
\]

We have seen that wellfoundedness suffices for recursiveness. In fact,
it is also necessary. While this equivalence is easy for polynomial
functors on the category of sets, it becomes remarkably involved in
more general settings, see Taylor \cite{Tay:prafm}.

For $F X = 1 + \El \times X \times X$, $A = \List$, $\alpha = \qsplit$,
we get this induction principle:
\[
\infer{P\, xs}{
  xs : \List
  &
  P\, \nil
  & 
  \infer*{P\, (\cons\,(x, xs'))}{
    x : \El 
    &
    xs' : \List
    &
    P\, (xs'|_{\leq x})
    &
    P\, (xs'|_{> x})
  }
}
\]

\paragraph{General case (2): Inductive graph relation} 

The original Bove-Capretta method separates determining the domain of
definition of a function from determining its values. Bove
\cite{Bov:anolfd} showed that this separation can be avoided.

For given $(A, \alpha)$, $(B, \beta)$, define a relation ${\dn}$
between $A$, $B$ \emph{inductively} by
\[
\infer{a \dn \be\, bs}{
  a : A & bs : F B
  &
  \al\, a\, \mathbin{(\tilde{F}\, {\dn})}\, bs
}
\]

Further, define a predicate $\Domg$ on $A$ by 
\[
\Domg\, a = \exists b : B.\, a \dn b
\]

It is straightforward to verify that $\forall a : A, b, b_* : B. a \dn b
\wedge a \dn b_* \to b = b_*$.  Moreover, it is also the case that
$\forall a : A.\, \Domg\, a \leftrightarrow \Dom\, a$.  So, $\Domg$
does not really depend on the given $(B, \beta)$!

From the last equivalence it is immediate that there is $f : A|_\Domg
\to B$ uniquely solving
\[
\xymatrix{
F(A|_\Domg) \ar[d]_{F f}
  & A|_\Domg \ar[l]_-{\al|_\Domg} \ar[d]^{f} \\
FB \ar[r]_-{\be} 
  & B
}
\]
And, if $\forall a : A.\, \Domg\, a$, which is the same as $A|_\Domg
\cong A$, then $f$ is a unique solution of the original equation.

As a matter of fact, recursiveness and wellfoundedness are equivalent
exactly because $\forall a : A.\, \Domg\, a \leftrightarrow \Dom\, a$.

For $F X = 1 + \El \times X \times X$, $A = \List$, $\alpha = \qsplit$,
$B = \List$, $\beta = \mathsf{concat}$, the
relation $\dn$ is defined inductively by
\[
\infer{\nil \dn \nil}{
}
\qquad
\infer{\cons\, (x, xs) \dn \mathsf{app}\, (ys_0, \cons(x, ys_1))}{
  x : \El
  &
  xs : \List
  & 
  xs|_{\leq x} \dn ys_0
  &
  xs|_{> x} \dn ys_1 
}
\]

\paragraph{Inductive domain and graph do not work for non-terminating
  productive definitions}

Unfortunately, for our $\dropevens$ example,
\[
\xymatrix@C=6em{
\El \times \Str \ar[d]_{1 + \El \times \dropevens}
  & \Str \ar[l]_-{\pair{\hd}{\tl \circ \tl}} \ar[d]^{\dropevens} \\
\El \times \Str \ar[r]_-{\red{\pair{\hd}{\tl}^{-1}}} 
  & \Str
}
\]
we get $\forall xs : \Str.\, \Dom\,  xs \equiv \bot$! Now, surely
there is a unique function from $0 \to \Str$. But this is
uninteresting! We would like to learn that there is a unique function
$\Str \to \Str$.

Intuitively, the reason why this equation has a unique solution lies
not in how a given argument is consumed but in how the corresponding
function value is produced. This is not a terminating but a productive
definition.

\paragraph{General case (3): Coinductive bisimilarity relation} The
concept of the domain of definedness can be dualized
\cite{CUV:corasg}. Besides partial solutions that are defined only on
a subset of the intended domain, it makes sense to consider ``fuzzy''
solutions that are defined everywhere but return values in a quotient
of the intended codomain.  But since the category of sets is not
self-dual, the theory dualizes only to a certain extent and various
mismatches arise.

For given $(B, \be)$, define a relation $\approx$ on $B$ \emph{coinductively}
by
\[
\infer{\exists bs, bs_* : FB.\, b = \beta\, bs \wedge b_* = \beta\, bs_* \wedge bs \mathbin{(\tilde{F}\,{\approx^*})} bs_*}{
  b, b_* : B
  &
  b \approx b_*
}
\]
(i.e., we take $\approx$ to be the largest/coarsest relation
validating this rule).

There need not necessarily be a function $f$ solving the equation
\[
\xymatrix{
F A \ar[d]_{F f}
  & A \ar[l]_-{\al} \ar[d]^{f} \\
F(B/_{\approx^*}) \ar[r]_-{\be/_{\approx^*}} 
  & B/_{\approx^*}
}
\]
but, if such a function exists, it can easily checked to be unique. (See
Capretta et al. \cite[Thm. 1]{CUV:corasg}.)

If $\forall b, b_* : B.\, b \approx b_* \to b = b_*$, which is the
same as $B/_{\approx^*} \cong B$ (where $B/_{\approx^*}$ is the
quotient of $B$ by the reflexive-transitive closure of $\approx$), we
say that $(B, \be)$ is \emph{antifounded}. If $(B, \be)$ is
antifounded, solutions to equation \ref{eq:rec} are the same as
solutions to the equation above, and thus unique.

For $F X = \El \times X \times X$, $B = \Str$, $\beta = \smerge$, the
relation $\approx$ is defined coinductively by
\[
  \infer{\begin{array}{l} \exists x : \El, xs_0, xs_1, xs_{0*}, xs_{1*} : \Str.\, \\ \hspace*{1cm}
     xs = \smerge(x, xs_0, xs_1) \wedge xs_* = \smerge(x, xs_{0*}, xs_{1*}) \wedge xs_0\, \approx\, xs_{0*} \wedge xs_1\, \approx\, xs_{1*}\end{array}}{
    xs, xs_* : \Str
    &
    xs\, \approx\, xs_*
  }
\]
It turns out that $\forall xs, xs' : \Str.\, xs \approx xs' \to xs =
xs'$. Based on this knowledge, we may conclude that solutions are
unique. (They do in fact exist as well for this example, but this has
to be verified separately.)

Solutions need not exist for antifounded algebras. E.g., for $F X =
X$, $B = \mathsf{Nat}$, $\beta = \mathsf{succ}$, we have that $(B,
\beta)$ is antifounded, but for $A$ any set and $\alpha =
\mathsf{id}_A$, the equation has the form $f\, a = \mathsf{succ}\, (f\
a)$ and has no solutions.

We have thus seen that antifoundedness of $(B, \be)$ does not
guarantee that it is corecursive. The converse also fails: not every
corecursive algebra $(B, \be)$ is antifounded \cite[Prop.
5]{CUV:corasg}.

However, for an antifounded algebra $(B, \be)$, we do get an
interesting coinduction principle on $B$: For any relation $R$ on $B$,
we have
\[
\infer{b = b_*}{
  b, b_* : B
  & 
  b\, R\, b_*
  &
  \infer*{\exists bs', bs'_* : F B.\, b' = \beta\, bs' \wedge b'_* = \beta\, bs'_* \wedge bs' \mathbin{(\tilde{F}\,{R^*})} bs'_*}{
    b', b'_* : B
    &
    b'\, R\, b'_*
  }
}
\]

For $F X = \El \times X \times X$, $B = \Str$, $\beta = \smerge$, we
get this coinduction principle:
\[
\infer{xs = xs_*}{
  xs, xs_* : \Str 
  &
  \hspace*{-1mm}
  xs\, R\, xs_*
  &
  \hspace*{-1mm}
  \infer*{\begin{array}{l} \exists x' : \El, xs'_0, xs'_1, xs'_{0*}, xs'_{1*} : \Str.\, \\ \hspace*{2mm}
     xs' = \smerge(x', xs'_0, xs'_1) \wedge xs'_* = \smerge(x', xs'_{0*}, xs'_{1x*}) \wedge xs'_0\, R\, xs'_{0*} \wedge xs'_1\, R\, xs'_{1*}\end{array}\hspace*{-1mm}}{
    xs', xs'_* : \Str
    &
    xs'\, R\, xs'_*
  }
}
\]

\paragraph{General case (4): Coinductive graph relation} Could one
also dualize the notion of the inductive graph? The answer is
positive. Differently from the case of the coinductive concept of
bisimilarity, this yields a criterion of unique solvability.

For given $(A, \alpha)$, $(B, \beta)$, define a relation ${\dnc}$
between $A$, $B$ \emph{coinductively} by
\[
\infer{\exists bs : F B.\, b = \be\, bs \wedge \al\, a\, \mathbin{(\tilde{F}\, {\dnc})}\, bs}{
  a : A & b : B
  &
  a \dnc b
}
\]

Define a predicate $\Domc$ on $A$ by 
\[
\Domc a = \exists b :
B.\, a \dnc b \]

Now we can construct $f : A|_\Domc \to B/_{\approx^*}$ that we can
prove to uniquely solve
\[
\xymatrix{
F(A|_\Domc) \ar[d]_{F f}
  & A|_\Domc \ar[l]_-{\al|_\Domc} \ar[d]^{f} \\
F(B/_{\approx^*}) \ar[r]_-{\be/_{\approx^*}} 
  & B/_{\approx^*}
}
\]

If both $\forall a : A.\, \Domc\, a$ and $\forall b, b_* : B.\, b
\approx b_* \to b = b_*$, which are the same as $A|_\Domc \cong A$
resp.\ $B/_{\approx^*} \cong B$, then $f$ uniquely solves also the
equation \ref{eq:rec}. Notice, however, that in this situation we have
obtained a unique solution only for the given $(A, \alpha)$: we have
\underline{not} established that $(B, \beta)$ is corecursive.

To formulate a further condition, we define a relation $\equiv$ on
$B$ by
\[ 
b \equiv b_* = \exists a : A.\, a \dnc b \wedge a \dnc b_*
\]
A unique solution to equation \ref{eq:rec} also exists if $\forall a :
A.\, \Domc\, a$ and $\forall b, b_* : B.\, b \equiv b_* \to b = b_*$.

This condition is weaker: while $\forall b, b_* : B.\, b \equiv b_*
\to b \approx b_*$, the converse is generally not true.

For $F X = \El \times X \times X$, $B = \Str$, $\beta = \smerge$
and any fixed $A$, $\alpha$, the relation $\dnc$ is defined
coinductively by
\[
\infer{\exists xs_0, xs_1 : \Str.\, xs = \smerge\, (\fst\, (\alpha\, a), xs_0, xs_1) \wedge \fst\, (\snd\, (\alpha\, a)) \dnc xs_0 \wedge \snd\, (\snd\, (\alpha\, a)) \dnc xs_1}{
  a : A 
  & 
  xs : \Str
  &
  a \dnc xs 
}
\]
It turns out that $\forall a : A.\, \Domc\, a$ 
no matter what $A$, $\alpha$ are.
So in this case we do have a unique solution $f$ for any $A, \alpha$,
i.e., $(\Str, \smerge)$ is corecursive.

\paragraph{Conclusion}

We have considered two flavors of partiality of a function: a function
may be defined only on a subset of the intended domain and the values
it returns may be underdetermined.

The Bove-Capretta method in its graph-based version scales meaningfully
to equations where unique solvability is not due to termination, but
productivity or a combination the two. But instead of one condition to
check by ad-hoc means, there are two in the general case.

The theory of corecursion/coinduction is not as clean as that of
recursion/induction---in particular, to admit coinduction is not the
same as to admit corecursion. We would like like to study the
coinductive graph approach further and to find out to what extent it
proves useful in actual programming practice. The main pragmatic issue
is the same as with Bove and Capretta's method: how to prove the
conditions.

\paragraph{Acknowledgments} This research was supported by Estonian
Science Foundation grant no.\ 6940 and the ERDF funded Estonian Centre
of Excellence in Computer Science, EXCS.


\begin{thebibliography}{99}
\providecommand{\urlalt}[2]{\href{#1}{#2}}
\providecommand{\doi}[1]{doi:\urlalt{http://dx.doi.org/#1}{#1}}

\bibitem{ALM:reccff} A. Ad\'amek, D. L\"ucke \& S. Milius (2007):
  \emph{Recursive coalgebras of finitary functors}. \textsl{Theor.
    Inform. and Appl.} 41(4), 447--462. \doi{10.1051/ita:2007028}

\bibitem{Bov:anolfd} A. Bove (2009): \emph{Another look at function
    definitions}.  In S.  Abramsky, M. Mislove \& C. Palamidessi,
  editors: \textsl{Proc.  of 25th Conf. on Mathematical Foundations of
    Programming Semantics, MFPS-XXV (Oxford, Apr. 2009)},
  \textsl{Electron. Notes in Theor. Comput. Sci.} 249, Elsevier,
  61--74. \doi{10.1016/j.entcs.2009.07.084}

\bibitem{BC:modgrt} A. Bove \& V. Capretta (2005): \emph{Modelling
    general recursion in type theory}. \textsl{Math. Struct. in
    Comput. Sci.}  15(4), 671--708. \doi{10.1017/s0960129505004822}

\bibitem{BC:typprf} A. Bove \& V. Capretta (2008): \emph{A type of
    partial recursive functions}. In O. A\"it Mohamed, C. Mu\~noz \&
  S.~Tahar, editors: \textsl{Proc. of 21st Int. Conf. on Theorem
    Proving in Higher Order Logics TPHOLs 2008 (Montreal, Aug. 2008},
  \textsl{Lect. Notes in Comput. Sci.} 5170, Springer, 102--117.
  \doi{10.1007/978-3-540-71067-7\_12}

\bibitem{BKS:parrit} A. Bove, A. Krauss \& M. Sozeau (2011):
  \emph{Partiality and recursion in interactive theorem provers: an
    overview}. Manuscript, submitted to \textsl{Math. Struct. in
    Comput. Sci.}.


\bibitem{CUV:reccc} V. Capretta, T. Uustalu \& V. Vene (2006):
  \emph{Recursive coalgebras from comonads}. \textsl{Inform.\ and
    Comput.} 204(4), 437--468. \doi{10.1016/j.ic.2005.08.005}

\bibitem{CUV:corasg} V. Capretta, T. Uustalu \& V. Vene (2009):
  \emph{Corecursive algebras: a study of general structured
    corecursion}. In M. V. M.  Oliveira \& J. Woodcock, editors:
  \textsl{Revised Selected Papers from 12th Brazilian Symp. on Formal
    Methods, SBMF 2009 (Gramado, Aug. 2009)}, \textsl{Lect.  Notes in
    Comput. Sci.} 5902, Springer, 84--100.
  \doi{10.1007/978-3-642-10452-7\_7}

\bibitem{Tay:prafm} P. Taylor (1999): \emph{Practical Foundations of
    Mathematics}, chapter VI.  Cambridge University Press.


\end{thebibliography}

\end{document}